\documentclass[a4paper]{article}
\usepackage{graphicx}

\begin{document}

\begin{center}
{\Large \bf
Empirical Parameterization of Nucleon-Nucleon Elastic Scattering Amplitude
at High Beam Momenta for Glauber Calculations and Monte Carlo Simulations
}
\end{center}

\begin{center}
V.~Uzhinsky\footnote{Laboratory of Information Technologies, JINR, Dubna, Russia.},
A.~Galoyan\footnote{Veksler and Baldin Laboratory of High Energy Physics, JINR, Dubna, Russia.},
Q.~Hu\footnote{Key Laboratory of High Precision Nuclear Spectroscopy and
               Center for Nuclear Matter Science, Institute of Modern Physics,
               Chinese Academy of Sciences, Lanzhou 730000, China.}$^,$\footnote{
               Institut f\"ur Kernphysik, Forschungszentrum J\"ulich, J\"ulich, 52425, Germany.},
J.~Ritman$^{4,}$\footnote{Ruhr-Universit\"at-Bochum, Bochum, 44780,
               Germany}$^,$\footnote{JARA-FAME, J\"ulich, 52425, Germany.},
H.~Xu$^{4}$
\end{center}

\begin{center}
\begin{minipage}{14cm}
A parameterization of the nucleon-nucleon elastic scattering amplitude is needed for future
experiments with nucleon and nuclear beams in the beam momentum range of 2 -- 50 GeV/c/nucleon.
There are many parameterizations of the amplitude at $P_{lab} >$ 25--50 GeV/c, and at
$P_{lab} \leq$ 5 GeV/c. Our paper is aimed to cover the range between 5 -- 50 GeV/c.

\vspace{3mm}
The amplitude is used in Glauber calculations of various cross sections and Monte Carlo
simulations of nucleon-nucleon scatterings. Usually, the differential nucleon-nucleon
elastic scattering cross sections are described by an exponential expression.
Corresponding experimental data on $pp$ interactions at $|t|>$ 0.005 (GeV/c)$^2$ and
$|t|\leq$ 0.125 (GeV/c)$^2$ have been fit. We propose formulae to approximate the beam momentum
dependence of these parameters in the momentum range considered. The same
was done for $np$ interactions at $|t|\leq$ 0.5 (GeV/c)$^2$. Expressions for the momentum
dependence of the total and elastic cross sections, and the ratio of real to imaginary
parts of the amplitude at zero momentum transfer are also given for $pp$ and $np$ collisions.
These results are sufficient for a first approximation of the Glauber calculations. For more exact
calculations we fit the data at $|t|>$ 0.005 (GeV/c)$^2$ without restrictions on the maximum
value of $|t|$ using an expression based on two coherent exponential. The parameters of the fits are
found for the beam momentum range 2 -- 50 GeV/c.

\end{minipage}
\end{center}

\setcounter{footnote}{0}
\section*{Introduction}
Parameterizations of nucleon-nucleon elastic scattering amplitude are widely used in many
Glauber calculations, for example, for studies of the structure of exotic nuclei and
the neutron skin of nuclei \cite{Alkhazov} -- \cite{ExoticA16}, calculations of
differential and total reaction cross sections for hadron-nucleus and
nucleus-nucleus interactions~\cite{ReactionX1,ReactionX2}, experimental research of high
energy nucleus-nucleus collisions at various impact parameters~\cite{RHIC1,RHIC2}, etc.
Considering a nucleon-nucleus interaction, it is sufficient for a first approximation to know
the amplitude at zero scattering angle, where the nucleon-nucleon interaction radius ($\sim 1$ fm)
is smaller than the nuclear size. The amplitude is connected with the total cross section and
the ratio of real to imaginary parts of the amplitude at small momentum transfer. Thus, the
parameterizations allow one to extract the total cross section, and the ration for calculations
of properties of hadron-nucleus scatterings.

For estimations of geometrical properties of inelastic hadron-nucleus and nucleus-nucleus reactions,
a so-called "inelastic nucleon-nucleon interaction profile" is used. The profile is also connected
with the amplitude. Very often simplified parameterizations are used for this profile. However,
modern experiments in high energy physics and planned experiments with exotic nuclei require
very accurate calculations. Thus, a good knowledge of the amplitude is very important for various
applications.

A standard parameterization of the spin averaged nucleon-nucleon elastic scattering amplitude is:
\begin{equation}
d\sigma/dt=\pi |F(t)|^2 = |A|^2 e^{B t},    \label{Eq1}
\end{equation}
\begin{equation}
\sigma^{tot}= 4\pi Im(F(0)),                \label{Eq2}
\end{equation}
\begin{equation}
A=\frac{\sigma^{tot}}{4\sqrt{\pi}}(i+\rho), \label{Eq3}
\end{equation}
$$\rho=Re(F(0))/Im(F(0)).$$

A set of its parameters ($A$, $B$, and sometimes $\sigma^{tot}$ and $\rho$) at various beam momenta
below 3 GeV/c \cite{Ray} is widely applied in calculations at low and intermediate energies (see also~\cite{EPAN}).
In principle, the parameters can be obtained by fitting the partial wave analysis results
\cite{PWA1} -- \cite{PWA7}, \cite{PWAweb} at momenta below 5 GeV/c as done in~\cite{Ray}.
Predictions of the partial wave analysis in a tabulated form are used in the PLUTO event
generator~\cite{PLUTO} and in the Geant4 toolkit~\cite{Geant4}.

At momenta higher than $\sim 20$ GeV/c, there are many parameterizations of the momentum
dependence of the total nucleon-nucleon cross section and $\rho$ including ones by
the Particle Data Group (PDG)~\cite{PDG2015}. Recently, formulae for values
of the slope parameter, $B$, were presented in~\cite{Okorokov}.

A large set of data on proton-proton differential cross sections, analyzing powers and the double
polarization parameter, $A_{NN}$, at proton beam momenta from 3~GeV/c to 50~GeV/c were analyzed in
\cite{Sibir} employing the Regge formalism. $\rho$, $\omega$, $f_2$, and $a_2$
trajectories and single-Pomeron exchange were considered. A complicated form of the reggeon
amplitudes prevents its simple application in Glauber calculations.

In Sec.~1, we present fits of differential proton-proton elastic scattering cross
sections by Eq.~\ref{Eq1} in the momentum range 2 -- 50 GeV/c. Approximate formulae for the
momentum dependence of the parameters are given. We also check the self-consistency
of the fitted $A$, $\sigma^{tot}$ and $\rho$ parameters. Here, we use the latest data
of COSY~\cite{COSY} to estimate $\rho$ in the momentum range 1.7 -- 3.6 GeV/c.

The simple exponential parameterization cannot describe the experimental data at large $|t|$
values, especially at $|t|\geq $1--1.5 (GeV/c)$^2$, where the slope of the differential
cross sections becomes smaller. Thus, integration of Eq.~\ref{Eq1} will lead to an
underestimation of elastic cross section, $\sigma^{el}$, and to an overestimation of the
inelastic one. A new expression for the differential cross section is needed.

A very simple and transparent parameterization of the elastic scattering amplitude
was proposed in Ref.~\cite{Kalbfleisch} with an analysis of antiproton-proton data at
$P_{lab}=$ 1.11, 1.33, and 1.52 GeV/c. A wider set of the $\bar pp$ data was considered in
references \cite{Crawley1,Crawley2,Crawley3,OurPbarP}. An analogous parameterization was
independently proposed in~\cite{TwoExpon}. The authors of that paper
analyzed only $pp$ experimental data at $P_{lab} =$12, 14.2, 19.2, 24, 29.7 GeV/c and at
$\sqrt{s_{pp}} =$ 53 GeV. However, the reduced $\chi^2/NDF$ values and the parameter errors were not given. Various extensions of the parameterization were proposed at higher momenta
\cite{Grau,Fagundes,Pancheri,Ster}. Its application to low momentum data is complicated
by a restricted range of $|t|$ in many cases. Thus, we simplified the model in Sec.~2
and included in our analysis data from the EDDA Collaboration~\cite{EDDA} ($E=$0.23 -- 2.59 GeV,
$\theta_{c.m.}=$30$^\circ$ -- 90$^\circ$). In addition to those data,
we added optical points that were calculated using $\sigma^{tot}$ and $\rho$.
This allowed us to clarify the behaviour of the fit parameters in the momentum
range 2.3 -- 3.8 GeV/c.

In Sec. 3 we turn to the analysis of $np$ elastic scattering data, and obtain analogous results.
A short conclusion is presented at the end of the paper.

\section{Standard parameterization of $pp$ elastic scattering data}
The amplitude of elastic proton-proton scattering must be symmetric when exchanging
$t\leftrightarrow u$, where $t$ and $u$ are Mandelstam variables. Thus, we
write the amplitude as $ F(t)=f(t)+f(u)$ and thus an expression to fit the differential
cross sections can be represented as:
\begin{equation}
d\sigma/dt=A^2\ (e^{B t/2}+e^{B u/2})^2. \label{Eq4}
\end{equation}

It is obvious that the parameterization can only be applied in a defined region of 4-momentum
transfer, $t$. The region must not include a region where the Coulomb interaction dominates.
Thus, we excluded experimental data with $|t|\ < \ 0.005$ (GeV/c)$^2$. The width of
the region must be sufficiently large, because any data set could be fit with
Eq.~\ref{Eq4} for a narrow range of $t$. We choose a maximum value of $|t|$ equal to 0.125
(GeV/c)$^2$ as selected in reference~\cite{Beznogikh}.
\begin{figure}[cbth]
\begin{center}
\includegraphics[width=150mm,height=40mm,clip]{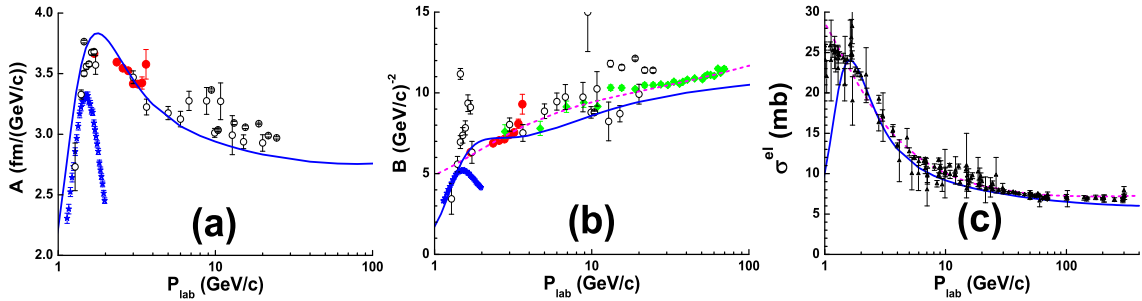}
\end{center}
\vspace{-5mm}
\caption{a, b) Fit parameters $A$ and $B$ as functions of the projectile momentum, $P_{lab}$.
Open circles are results of the fit to the data \protect{\cite{LIYAF} -- \cite{AllabyW}}.
Filled circles are the results for the latest COSY data \protect{\cite{COSY}}. Stars present results of the fit
to the EDDA data \protect{\cite{EDDA}}. Diamonds are data from reference \protect{\cite{Beznogikh}} (see Fig.~1b).
Dashed (magenta) curves are approximations of the dependencies. Descriptions of the solid curves are given in the text.
c) Elastic cross section as a function of $P_{lab}$. The data points are
from the PDG data-base
\protect{\cite{PDG2015}}.}
\label{Fig1}
\end{figure}

Results\footnote{
$\chi^2/NDF=938/642\simeq 1.46$. Without fitting the EDDA data \cite{EDDA} and
small angle data at $P_{lab}=$ 9.43 and 18.9 GeV/c, $\chi^2/NDF=533/579\simeq 0.92$.
NDF $=$ Number of Degree of Freedom.}$^)$
of fits to the data (various symbols) within these $t$-ranges are presented in Fig.~1.
The parameter $A$ is observed to grow from $P_{lab}\simeq 1$ GeV/c, reaches a maximum at
$P_{lab}\simeq 1.7$ GeV/c and then decreases at higher momentum (see Fig.~1a).
Results for the COSY data are in a good agreement with the other data.
A fit to only the EDDA data \cite{EDDA} gives acceptable results at $P_{lab} \leq 1.7$ GeV/c.
At higher $P_{lab}$, the parameter $A$ for these data decreases, reflecting the fact that
small angle data are absent in that measurement.

The parameter $B$ grows for $P_{lab} \geq 1$ GeV/c, and then sharply decreases above
$P_{lab} \simeq 1.7$ GeV/c and continues with a smooth growth for higher momentum
(see Fig.~1b). Our fit results are in agreement with previous ones \cite{Beznogikh} at
$P_{lab} > 10$ GeV/c. The COSY data clarify the behaviour at $P_{lab}=$ 1.7 -- 3.6
GeV/c. The EDDA data cannot be used to determine these parameters for
$P_{lab} > 1.7$ GeV/c, because they do not measure to sufficiently small momentum transfer.

If the fit and the data on $\sigma^{tot}_{pp}$ and $\rho_{pp}$ are
self-consistent, then the parameter $A$ must be connected with $\sigma^{tot}_{pp}$ and $\rho_{pp}$ according
to Eq.~\ref{Eq3}. To check this we need parameterizations of $\sigma^{tot}_{pp}$ and $\rho_{pp}$,
because the momentum range of these data does not coincide with the range where
$\sigma^{tot}_{pp}$ and $\rho_{pp}$ have been determined. Extending the PDG
parameterizations \cite{PDG2015}, we approximate $\sigma^{tot}_{pp}$ and
$\rho_{pp}$ by the following formulae:
\begin{equation}
\sigma^{tot}_{pp}=\sigma^{tot}_{PDG}+\frac{7200}{(s/s_0)^{3.5}}-\frac{32}{(s/s_0-4)^2+0.45}
\ \ \ [{\rm mb}],                                                                    \label{Eq5}
\end{equation}
\begin{equation}
\rho_{pp} = \rho_{PDG}+\frac{1.5}{(s/s_0)} - \frac{3}{(s/s_0)^2}, \ \ \ s_0=1\ {\rm GeV}^2,  \label{Eq6}
\end{equation}
where $s$ is the center-of-mass energy squared ($s=2m_N^2+2m_N(E+m_N)$) in GeV$^2$,
$m_N$ is the nucleon mass (0.938 $GeV/c^2$), and
\begin{equation}
\sigma^{tot}_{PDG}=H\ \ln^2(s/s_M) + P + R_1 (s/s_M)^{-\eta_1} - R_2 (s/s_M)^{-\eta_a2}, \label{Eq7}
\end{equation}
\begin{equation}
\rho_{PDG}=\frac{1}{\sigma^{tot}_{PDG}}\left[ \pi H \ln(s/s_M) -
           R_1 (s/s_M)^{-\eta_1} \tan\left(\frac{\eta_1 \pi}{2}\right) -
           R_2 (s/s_M)^{-\eta_2} \cot\left(\frac{\eta_2 \pi}{2}\right)\right],        \label{Eq8}
\end{equation}
$$s_M=(2m_N+M)^2,\ \ \ M=2.076\ [{\rm GeV}], \ \ \ H=0.2838\ [{\rm mb}],\ \ \ P=33.73\ [{\rm mb}],$$
$$R_1=13.67\ [{\rm mb}],\ \ \ \eta_1=0.412,\ \ \ R_2=7.77\ [{\rm mb}],\ \ \ \eta_2=0.5626.$$

Experimental data on $\sigma^{tot}_{pp}$ and $\rho_{pp}$ from the PDG data-base are presented in Fig.~2 together with our parameterizations. The PDG parameters were
determined at $\sqrt{s} \geq 7$ GeV. Direct extrapolations of the PDG parameterizations
below 7 GeV are shown in Fig.~2 by dashed lines. It is obvious, that they do not
describe the data in the low momentum domain. Thus, we include additional terms in
our approximations.

The forms of the additional terms are mainly motivated by the reggeon phenomenology.
According to the phenomenology, a yield of a non-vacuum reggeon exchange to the
elastic scattering amplitude is proportional to $1/s^n$ at high momenta, where
$n$ can be $\sim$ 0.5, 1, 1.5, 2 and so on for various reggeons. The PDG parameterization
of the total cross sections include only two effective non-vacuum reggeon exchanges --
terms $R_1 (s/s_M)^{-\eta_1}$ and $R_2 (s/s_M)^{-\eta_2}$.

We did not extend our parameterization for $\rho_{pp}$ below 1 GeV/c,
because the behavior of the $\rho_{pp}$ data is unclear for these momenta.

The latest COSY data~\cite{COSY} helped to estimate $\rho_{pp}$ in the momentum
range 1.7 -- 3.6 GeV/c. According to Eq.~\ref{Eq3}, $\rho_{pp}=-\sqrt{16\pi \ (A/\sigma^{tot}_{pp})^2-1}$.
We calculated $\rho_{pp}$ using the fit results for $A$ and the approximation for $\sigma^{tot}_{pp}$
according to Eq.~\ref{Eq5}. The calculations are presented in Fig.~2 by stars. They clarify
the behavior of $\rho_{pp}$ at momenta below 3.8 GeV/c. After that, we determined parameters
of the approximation in Eq.~\ref{Eq6}.

Having the approximations in Eqs.~\ref{Eq5} and \ref{Eq6}, we can now investigate
Eq.~\ref{Eq3} in the momentum range under study. The estimated values
of $A$ are presented by the solid line in Fig.~1a. As seen, the estimations
are in reasonably good agreement with the fit results for $P_{lab} \leq 8$ GeV/c (i.e. the estimations
do not deviate from experimental values by more than experimental error bars).
The estimations are lower than the fit results by about 5 -- 10 \% for $P_{lab}=$ 8 -- 25 GeV/c.
This can be connected with an overestimation of $\rho_{pp}$ in the momentum region:
$\rho_{pp} \simeq -0.35$ according to the Eq. \ref{Eq6}. It is sufficient to reach an
agreement between the estimations and the fit results for $A$ to decrease $\rho_{pp}$
down to -0.45 as it was for the COSY momenta. Thus, we believe that $\rho_{pp}$ can
have a non-trivial behavior in this momentum region. It could be a main reason
why the high momentum approximation cannot be extended in the low momentum domain.

Of course, that  disagreement could result from an over-simplified
exponential expression. We evaluate this hypothesis in the next section.

\begin{figure}[cbth]
\begin{center}
\includegraphics[width=150mm,height=120mm,clip]{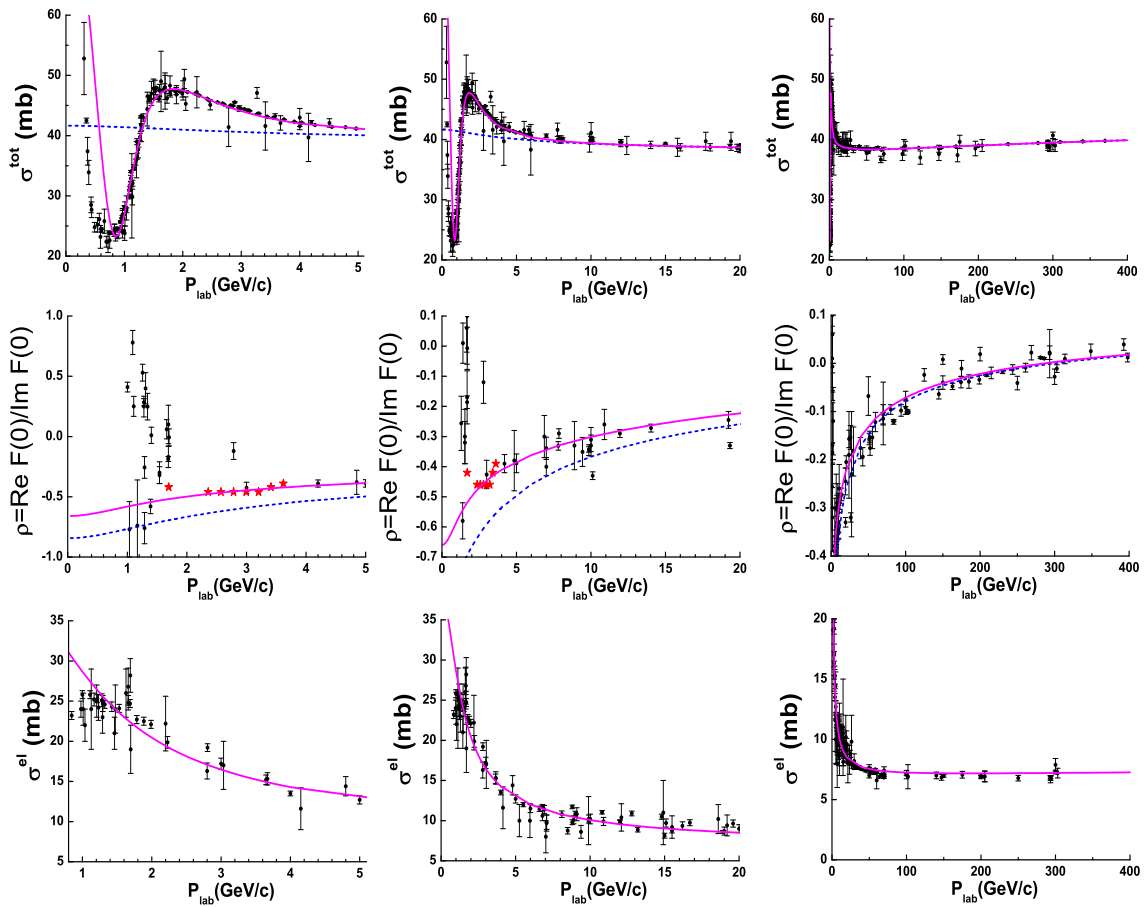}  
\end{center}
\vspace{-5mm}
\caption{$\sigma^{tot}_{pp}$, $\sigma^{el}_{pp}$ and $\rho_{pp}$ as functions of $P_{lab}$.
Points are data from PDG data-base \protect{\cite{PDG2015}}. Dashed lines are
extrapolations of the PDG parameterizations. The solid lines are our approximations.}
\label{Fig2}
\end{figure}

It is useful to have an approximation of the momentum dependence of the
parameter $B$ for the Glauber calculations of the differential cross
sections. We parameterize the fit results for $B$ by the following formula:
\begin{equation}
B=1.9\ \ln(s/s_0) + \frac{27}{\sqrt{s/s_0}} - \frac{47}{(s/s_0)}\ \ \ [({\rm GeV/c})^{-2}].     \label{Eq9}
\end{equation}
Calculations using this formula are shown as a dashed line in Fig.~1b.

Using the approximated values for $A$ and $B$, we calculate the elastic
cross sections ($\sigma^{el}_{pp}=|A|^2/B$) presented in Fig.~1c by a solid
line. As expected, the calculations somewhat underestimate the cross
sections. To show this, we fit the elastic cross sections with the following expression:
\begin{equation}
\sigma^{el}_{pp}=0.18\ \sigma^{tot}_{PDG}+\frac{60}{(s/s_0)}+\frac{600}{(s/s_0)^3} \ \ \ [{\rm mb}],\label{Eq10}
\end{equation}
and plot it in Fig.~1c as a dashed line. From the other hand, the approximations
for $\sigma^{tot}_{pp}$, $\sigma^{el}_{pp}$ and $\rho_{pp}$ can be used to calculate $B$ as
\begin{equation}
B=\frac{(\sigma^{tot}_{pp})^2 (1+\rho^2_{pp})}{16\ \pi\ \sigma^{el}_{pp}}\ 2.568\ \ \
[({\rm GeV/c})^{-2}],\label{Eq11}
\end{equation}
if $\sigma^{tot}_{pp}$ and $\sigma^{el}_{pp}$ are given in millibarn. The calculated values of $B$ are
shown as a solid line in Fig.~1b. As seen, the calculations also underestimate the fit results.
Note, that Eq.~\ref{Eq11} allows one a correct reproduction of the inelastic cross sections.

\section{Two exponential parameterization of $pp$ data}
The two exponential parameterization is considered in order to describe experimental data over a wide range of $t$.
\begin{equation}
\frac{d\sigma}{dt}=|A_1\ e^{B_1t/2}\ + \ A_2e^{i\phi}\ e^{B_2t/2}|^2 ,\label{Eq12}
\end{equation}
where $A_1$, $B_1$, $A_2$, $B_2$ and $\phi$ are real numbers.
It was proposed in Ref.~\cite{Kalbfleisch} for an analysis of
anti-proton-proton elastic scattering data and was also applied to describe
a wide set of $\bar{p}p$ data in the momentum range 1 -- 15 GeV/c in~\cite{Crawley1,Crawley2,Crawley3}.

This idea was independently proposed by R.J.N.~Phillips and V.D.~Barger~\cite{TwoExpon} in 1973.
They analyzed only $pp$ experimental data at $P_{lab} =$12, 14.2, 19.2,
24, 29.7 GeV/c and at $\sqrt{s_{pp}} =$ 53~GeV for the range $0.15<|t|<5$~(GeV/c)$^2$.

This paramterization has been used to fit the available experimental data
within the range $0.005<|t|<5$~(GeV/c)$^2$. Results of the fit are presented in
Tab.~1. The values of the parameters for the two restrictions on $t$
($0.15<|t|$ \cite{TwoExpon} and $0.005<|t|$ as before) differ by no more
than 10 \%. The typical difference is about 5 \%. Our restriction allowed additional
data sets to be included in the fitting procedure.
\begin{table}[cbth]
\begin{center}
\caption{Results of five-parameter fits with Eq. \ref{Eq12} for $pp$ interactions.}
\begin{tabular}{|c|c|c|c|c|c|c|}
\hline
$P_{lab}$ &    $A_1$   &  $B_1$         &    $A_2$   & $B_2$          & $\phi$  & $\chi^2/NDF$\\ \hline
[GeV/c]   &[fm/(GeV/c)]&[(GeV/c)$^{-2}$]&[fm/(GeV/c)]&[(GeV/c)$^{-2}$]& [rad]   &             \\ \hline
   10.0  &  2.48 $\pm$  0.08 &  7.85 $\pm$ 0.28 & 0.0790 $\pm$ 0.0190 &  1.20 $\pm$ 0.26 &  1.59  $\pm$ 0.24 & 0.93\\ \hline
   12.0  &  2.76 $\pm$  0.08 &  7.95 $\pm$ 0.20 & 0.0772 $\pm$ 0.0079 &  1.28 $\pm$ 0.09 &  1.73  $\pm$ 0.13 & 0.42\\ \hline
   14.2  &  2.37 $\pm$  0.11 &  7.77 $\pm$ 0.23 & 0.0948 $\pm$ 0.0066 &  1.67 $\pm$ 0.05 &  2.13  $\pm$ 0.08 & 0.23\\ \hline
   19.2  &  2.67 $\pm$  0.05 &  8.09 $\pm$ 0.09 & 0.7780 $\pm$ 0.0021 &  1.76 $\pm$ 0.02 &  2.11  $\pm$ 0.04 & 1.33\\ \hline
   20.0  &  2.77 $\pm$  0.02 &  8.51 $\pm$ 0.08 & 0.0682 $\pm$ 0.0064 &  1.72 $\pm$ 0.10 &  1.88  $\pm$ 0.07 & 1.56\\ \hline
   21.12 &  2.52 $\pm$  0.04 &  8.18 $\pm$ 0.11 & 0.0466 $\pm$ 0.0076 &  1.26 $\pm$ 0.18 &  1.93  $\pm$ 0.09 & 2.92\\ \hline
   24.0  &  2.51 $\pm$  0.05 &  7.98 $\pm$ 0.10 & 0.0707 $\pm$ 0.0030 &  1.84 $\pm$ 0.03 &  2.22  $\pm$ 0.05 & 0.62\\ \hline
   29.7  &  2.56 $\pm$  0.03 &  8.59 $\pm$ 0.13 & 0.0451 $\pm$ 0.0111 &  1.58 $\pm$ 0.26 &  2.09  $\pm$ 0.11 & 1.29\\ \hline
   50    &  2.70 $\pm$  0.01 &  9.60 $\pm$ 0.06 & 0.0189 $\pm$ 0.0023 &  1.60 $\pm$ 0.10 &  1.75  $\pm$ 0.10 & 2.59\\ \hline
   200   &  2.48 $\pm$  0.01 &  9.55 $\pm$ 0.03 & 0.0055 $\pm$ 0.0006 &  1.29 $\pm$ 0.07 &  2.33  $\pm$ 0.05 & 6.56\\ \hline
   293   &  1.17 $\pm$  0.08 &  7.67 $\pm$ 0.16 & 0.0207 $\pm$ 0.0024 &  2.04 $\pm$ 0.09 &  2.97  $\pm$ 0.02 & 0.90\\ \hline
   501   &  2.42 $\pm$  0.01 &  9.56 $\pm$ 0.19 & 0.0895 $\pm$ 0.0002 &  1.53 $\pm$ 0.02 &  2.95  $\pm$ 0.02 & 5.39\\ \hline
\end{tabular}
\end{center}
\end{table}
The parameters have been constrained as in Ref.~\cite{Crawley3}, because Eq.~\ref{Eq12}
contains 5 parameters that are often strongly correlated.
Due to the correlation the fit does not always converge.
Thus, we have introduced a constraint to reduce the number of free
parameters. Three of these constraints were considered in Ref.~\cite{Crawley3}:
$B_1=10$ (GeV/c)$^{-2}$, $B_2=B_1/3$ and $\phi=2.793$. According to Tab.~1,
$B_1$ varies by about $\pm$ 8 \%, $B_2/B_1$ by about $\pm$ 42 \%, and
$\phi$ by about $\pm$ 16 \%. Considering the fit results with the simple
exponential expression (see Fig.~1b), it is difficult to assume that $B_1$ is
a constant in the momentum range studied. $B_2/B_1$  also varies too strongly. Thus,
we assume that $\phi$ is approximately constant and set it to an average value
from Tab.~1, $\phi =$1.94 rad, at $P_{lab}\leq$ 50 GeV/c. We repeat the fit with a constant value of
$\phi$. Results of the fit are given in Tab.~2 and  presented
by solid points in Fig.~3.
\begin{table}[cbth]
\begin{center}
\caption{Results of fits with Eq. \ref{Eq12} for $pp$ interactions at $\phi =$1.94 rad.}
\begin{tabular}{|c|c|c|c|c|c|}
\hline
$P_{lab}$ &    $A_1$   &  $B_1$         &    $A_2$   & $B_2$          & $\chi^2/NDF$\\ \hline
[GeV/c]   &[fm/(GeV/c)]&[(GeV/c)$^{-2}$]&[fm/(GeV/c)]&[(GeV/c)$^{-2}$]&                        \\ \hline
    5.5  & 13.15 $\pm$ 0.06 &   4.21 $\pm$ 0.12 & 0.208 $\pm$ 0.010 & 1.30 $\pm$ 0.04 & 2.13\\ \hline
   10.0  & 24.12 $\pm$ 0.06 &   7.43 $\pm$ 0.10 & 0.117 $\pm$ 0.009 & 1.59 $\pm$ 0.10 & 1.01\\ \hline
   12.0  & 26.91 $\pm$ 0.06 &   7.66 $\pm$ 0.09 & 0.090 $\pm$ 0.004 & 1.41 $\pm$ 0.05 & 0.53\\ \hline
   14.2  & 25.34 $\pm$ 0.10 &   8.22 $\pm$ 0.14 & 0.084 $\pm$ 0.003 & 1.59 $\pm$ 0.03 & 0.46\\ \hline
   18.4  & 26.76 $\pm$ 0.13 &   8.55 $\pm$ 0.23 & 0.239 $\pm$ 0.140 & 3.54 $\pm$ 0.96 & 0.87 \\ \hline
   19.2  & 27.60 $\pm$ 0.05 &   8.40 $\pm$ 0.06 & 0.071 $\pm$ 0.001 & 1.72 $\pm$ 0.01 & 1.90\\ \hline
   20.0  & 27.61 $\pm$ 0.02 &   8.44 $\pm$ 0.03 & 0.072 $\pm$ 0.005 & 1.76 $\pm$ 0.09 & 1.53\\ \hline
   21.12 & 25.12 $\pm$ 0.03 &   8.16 $\pm$ 0.05 & 0.048 $\pm$ 0.003 & 1.29 $\pm$ 0.07 & 2.70 \\ \hline
   24.0  & 25.99 $\pm$ 0.04 &   8.40 $\pm$ 0.07 & 0.060 $\pm$ 0.002 & 1.75 $\pm$ 0.02 & 1.53\\ \hline
   29.7  & 25.76 $\pm$ 0.03 &   8.73 $\pm$ 0.05 & 0.035 $\pm$ 0.005 & 1.34 $\pm$ 0.17 & 1.30\\ \hline
   50.0  & 26.86 $\pm$ 0.01 &   9.49 $\pm$ 0.03 & 0.022 $\pm$ 0.002 & 1.70 $\pm$ 0.08 & 2.64\\ \hline
\end{tabular}
\end{center}
\end{table}

As seen in Fig.~3, the results for the data at $P_{lab}=$5.5~GeV/c
\cite{Kammerud} and 10~GeV/c \cite{Allaby} are far off the other results.
This is because the data at 5.5 GeV/c have no points for $|t| \leq$ 0.66 (GeV/c)$^2$ (see Fig.~4a). Thus,
the parameters $A_1$ and $B_1$ cannot be determined correctly.
The experimental data at $P_{lab}=$5.5~GeV/c
and data at $P_{lab}=$5.0~GeV/c \cite{Ambats} are plotted in Fig.~4a. The last data have
points at small $|t|$ but do not have enough points at large $|t|$.
\begin{figure}[cbth]
\begin{center}
\includegraphics[width=150mm,height=70mm,clip]{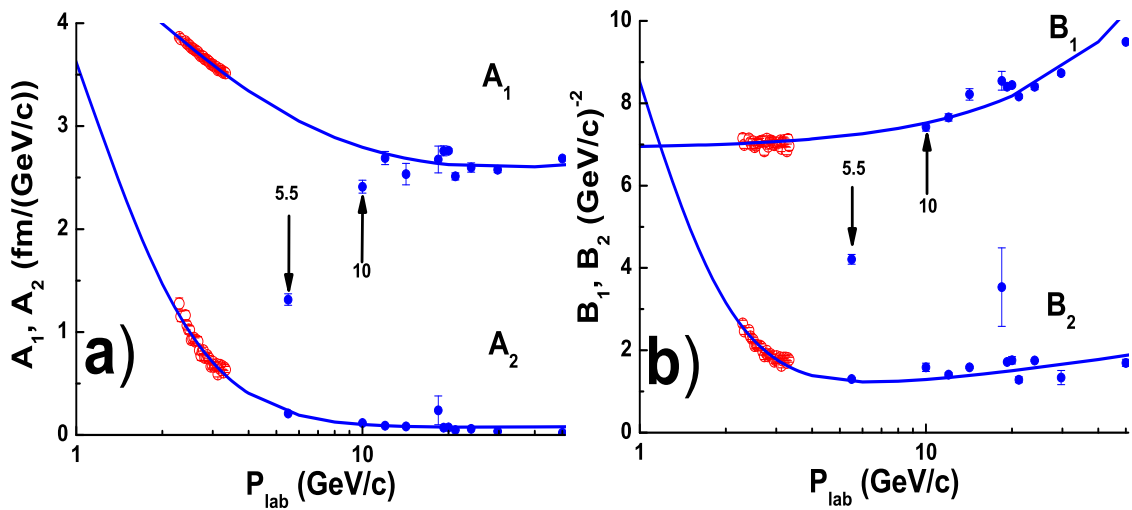}
\end{center}
\vspace{-5mm}
\caption{Fitted parameters $A_1$, $A_2$, $B_1$ and $B_2$ as functions of projectile momentum.
Solid points (blue) are results of the fitting of the data \protect{\cite{LIYAF}
-- \cite{Amaldi}} with constant $\phi$. Open points (red) are results for the
EDDA data \protect{\cite{EDDA}}. Arrows mark the results for $A_1$ and $B_1$ at 5.5 and 10 GeV/c.
Solid lines are approximations (see below).}
\label{Fig3}
\end{figure}

A more complicated situation occurs at $P_{lab}=$10~GeV/c
\cite{Allaby}. We compare those data with data at 9.9~GeV/c
\cite{Edelstein} in Fig.~4b. As seen, the data at 10~GeV/c
 do not have sufficient points at low $|t|$. In addition,
the points at 10~GeV/c fluctuate more strongly than the data
at 9.9~GeV/c. All of these reflect on the fit results.

To clarify the parameter's behaviour at $P_{lab}<$3~GeV/c, we have
included the EDDA data into the fit. The fit does not converge
because the data only contain differential cross section values at large
scattering angles. To overcome the problem, we
added values of $d\sigma/dt$ at $t=0$ to the data. These values
were calculated using $\sigma^{tot}_{pp}$ and $\rho_{pp}$ according to
Eq.~\ref{Eq5} and \ref{Eq6}. Errors of the values were set 0.5 \%.
Results of the fit are presented by open points in Fig.~3.

\begin{figure}[cbth]
\begin{center}
\includegraphics[width=150mm,height=60mm,clip]{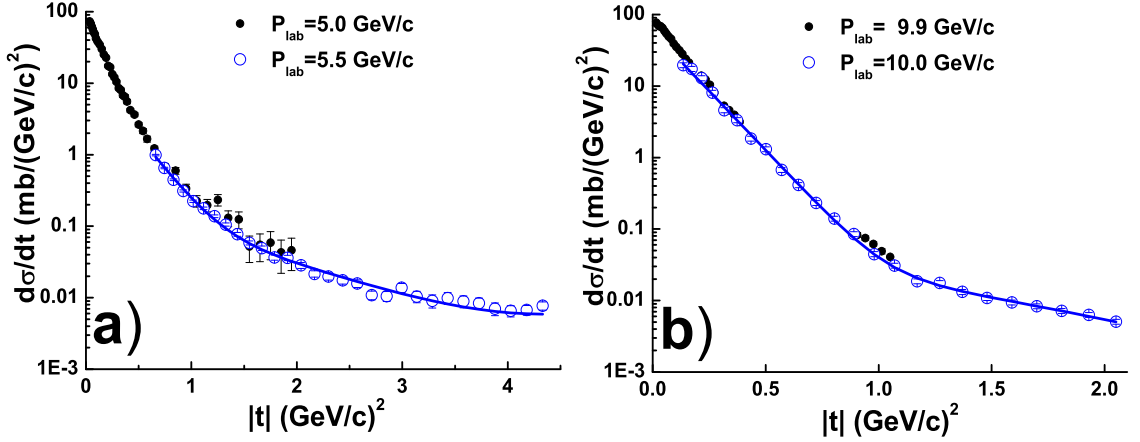}
\end{center}
\vspace{-5mm}
\caption{Differential cross sections at various momentum transfers.
The solid points are reference data \protect{\cite{Ambats,Edelstein}}, and
the open points have been fit \protect{\cite{Ambats,Kammerud,Allaby}}.
The blue solid lines are the results of the fit.}
\label{Fig4}
\end{figure}

Future calculations require approximations of the momentum dependence of $A_1$, $A_2$, $B_1$ and
$B_2$. A fit to the high momentum data
($\sqrt{s}>$ 23~GeV) to Eq.~12 has been done in Refs. \cite{Fagundes,Ster}, and they find
that $A_1$ and $B_1$ smoothly grow in the range
$23<\sqrt{s}\leq 7000$ GeV. $A_2$ and $B_2$ have more complicated behaviour,
however the following simple behaviours were proposed in Ref.\cite{Ster}:
\begin{equation}
A_1=a\ (s/s_0)^{-\epsilon_1}+b\ (s/s_0)^{\epsilon_2}, \ \ \
A_2=c\ (s/s_0)^{-\epsilon_3}+d\ (s/s_0)^{\epsilon_4}, \label{Eq13}
\end{equation}
\begin{equation}
B_1=b_0+b_1 \ln{(s/s_0)}, \ \ \ B_2=b_2+b_3 \ln{(s/s_0)}.                                 \label{Eq14}
\end{equation}
We were not able to select the parameters of Eq. \ref{Eq14} in the momentum range studied. Thus, we
changed the expressions for $B_1$ and $B_2$ to:
\begin{equation}
B_1=b_0\ (s/s_0)^{-\epsilon_5}+b_1\ (s/s_0)^{\epsilon_6}, \ \ \
B_2=b_2\ (s/s_0)^{-\epsilon_7}+b_3\ (s/s_0)^{\epsilon_8}. \label{Eq15}
\end{equation}

A careful selection of the parameters of the expressions resulted in:
\begin{equation}
a=10.6\ [{\rm fm}/({\rm GeV/c})],\ \ \epsilon_1=0.9,\ \
b=1.55\ [{\rm fm}/({\rm GeV/c})],\ \ \epsilon_2=0.1\  \label{Eq16}
\end{equation}
\begin{equation}
c=290\ [{\rm fm}/({\rm GeV/c})],\ \ \epsilon_3=3,\ \ \ \
d=0.05\ [{\rm fm}/({\rm GeV/c})],\ \ \epsilon_4=0.1\ \label{Eq17}
\end{equation}
\begin{equation}
b_0=6.8\ [({\rm GeV/c})^{-2}],\ \ \ \epsilon_5=0.,\ \ \ \
b_1=0.035\ [({\rm GeV/c})^{-2}],\ \ \epsilon_6=1.0\  \label{Eq18}
\end{equation}
\begin{equation}
b_2=2700\ [({\rm GeV/c})^{-2}],\ \ \epsilon_7=4,\ \ \ \ \ \
b_3=0.6\ [({\rm GeV/c})^{-2}],\ \ \ \ \epsilon_8=0.25\ \label{Eq19}
\end{equation}

Having the approximations, we can corroborate the self-consistency of the fit.
Neglecting the $t\leftrightarrow u$ symmetry at sufficiently high momenta,
a general form of the two exponential parameterization can be represented as
\begin{equation}
F(t)= e^{i\phi_0}\ [A_1\ e^{B_1t/2}\ + \ A_2e^{i\phi}\ e^{B_2t/2}]. \label{Eq20}
\end{equation}
Thus,
\begin{equation}
\sigma^{tot}=4\pi\ Im(F(0))=4\pi\ \{\sin{(\phi_0)}\ [A_1+A_2 \cos{(\phi)}]+A_2 \cos{(\phi_0)} \sin{(\phi})\}
\ \frac{1.974}{\sqrt{\pi}}\ \ \ [{\rm mb}], \label{Eq21}
\end{equation}
\begin{equation}
\phi_0=\pi\ + \arctan\left\{
\frac{[A_1+A_2 \cos{(\phi)}]-\rho A_2 \sin{(\phi})}{\rho [A_1+A_2 \cos{(\phi)}]+A_2 \sin{(\phi})}\right\}. \label{Eq22}
\end{equation}
Using $\rho_{pp}$ given by Eq.~\ref{Eq6} and approximations of the parameters, we have
calculated $\sigma^{tot}_{pp}$ and confirmed that the obtained values coincide with ones
predicted by Eq.~\ref{Eq5} to the level of $\pm$ 5 \%. It is a typical precision
of our estimations. New, more accurate experimental data
on $pp$ elastic scattering are needed in order to increase the precision.

\section{Parameterization of the  $np$ elastic scattering amplitude}
\begin{figure}[cbth]
\begin{center}
\includegraphics[width=150mm,height=140mm,clip]{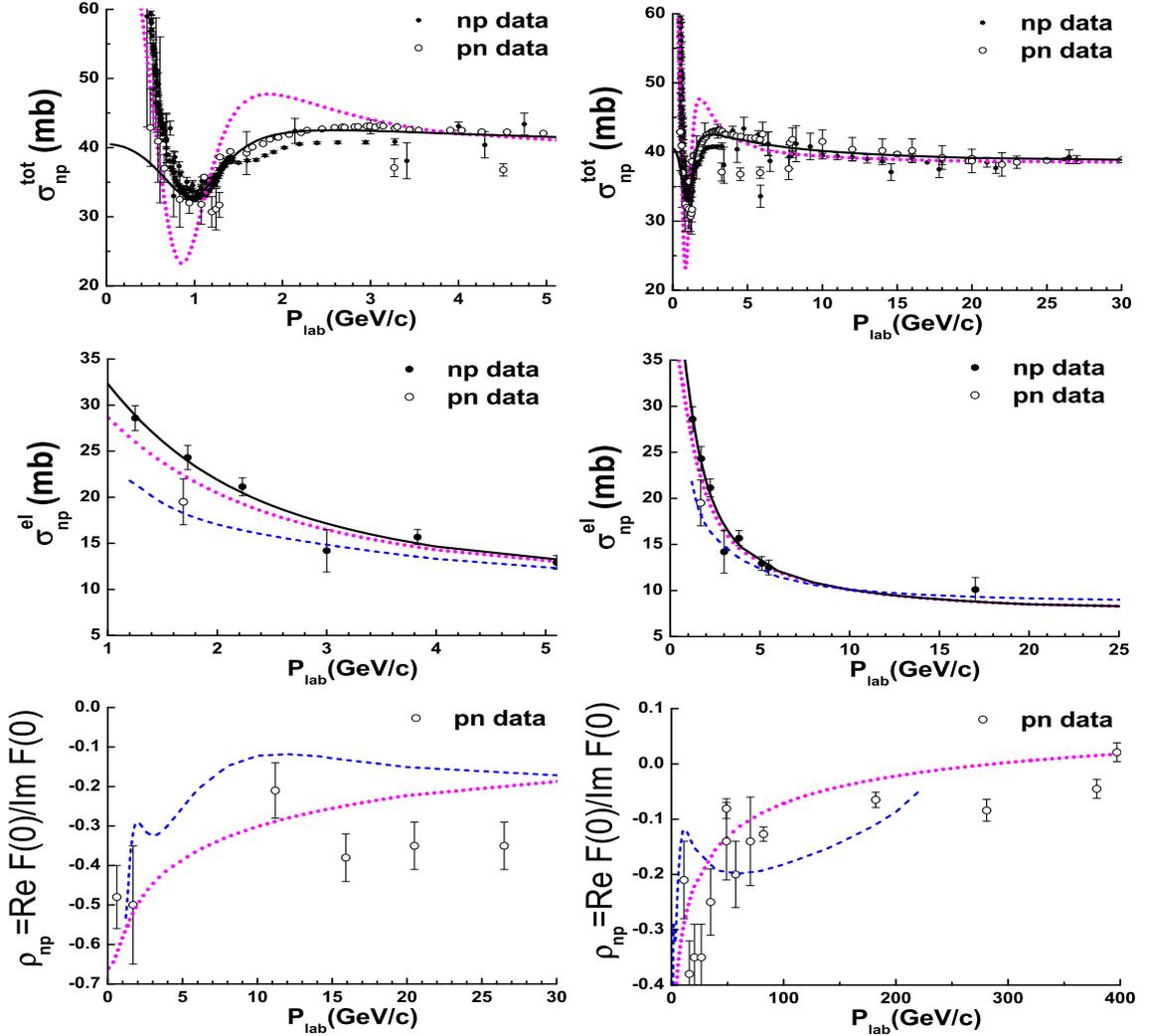}   
\end{center}
\vspace{-5mm}
\caption{$\sigma^{tot}$, $\sigma^{el}$ and $\rho$ of $pn$ and $np$ interactions as functions of $P_{lab}$.
The data points are from \protect{\cite{PDG2015}}. Solid (black) lines are approximations
for $pn$ and $np$ interactions. Dotted (magenta) lines are our approximations for $pp$ interactions.
Dashed (blue) lines are calculations (see the text).}
\label{Fig5}
\end{figure}

General properties of the $np$ interaction -- $\sigma^{tot}_{np}$, $\sigma^{el}_{np}$ and $\rho_{np}$
are presented in Fig.~\ref{Fig5} together with our approximations for $pp$ collisions.
The total $np$ cross section is lower than the $pp$ cross section for
$P_{lab}=$ 1.2 -- 3.4~GeV/c. They aproach each other at higher momenta.
The data on total cross sections of $pn$ and $np$ interactions are different in the region
$P_{lab}=$ 1.2 -- 2.2~GeV/c. It is a consequence of the different methods applied for the
measurements and the complicated structure of the differential cross section. We
will not consider the difference in detail, but we will assume that the
$pn$ data are more precise than $np$ ones.

As also seen in the figure, the total elastic scattering cross section for $np$
interactions is larger than the analogous data for $pp$ collisions at
$P_{lab}\leq$ 3.3~GeV/c. There is not sufficient data on $\rho=Re(F(0))/Im(F(0))$ for $np$ interactions
to draw a solid conclusion. Nevertheless, since they generally agree with
$pp$ data, we have assumed that $\rho_{np}=\rho_{pp}$.

Because the properties of $np$ interactions are similar to those of $pp$ interactions, we approximate
the momentum dependence of both $\sigma^{tot}_{np}$ and $\sigma^{el}_{np}$ by expressions analogous to Eqs.
\ref{Eq5} and \ref{Eq10} with an additional terms.
\begin{equation}
\sigma^{tot}_{np}=\sigma^{tot}_{PDG}+\frac{18}{(s/s_0)}-\frac{6.4}{(s/s_0-4.25)^2+0.5}
\ \ \ [{\rm mb}],                                                                    \label{Eq23}
\end{equation}
\begin{equation}
\sigma^{el}_{np}=0.18\ \sigma^{tot}_{PDG}+\frac{60}{(s/s_0)}+\frac{900}{(s/s_0)^3} \ \ \ [{\rm mb}],\label{Eq24}
\end{equation}
where $\sigma^{tot}_{PDG}$ is given by Eq.~\ref{Eq7}.

\begin{figure}[cbth]
\begin{center}
\includegraphics[width=150mm,height=60mm,clip]{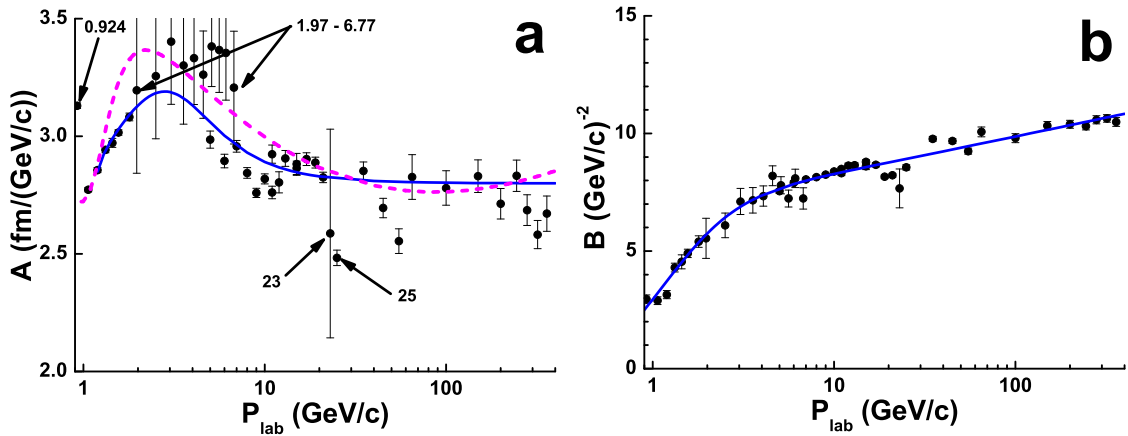}
\end{center}
\vspace{-5mm}
\caption{Fit results of Eq.~\ref{Eq1} to $np$ differential elastic scattering cross sections
         \protect{\cite{Terrien} -- \cite{Arefiev}} at $|t|<$ 0.5~(GeV/c)$^2$. Points
         are the fit results. Solid lines are approximations (Eqs. \protect{\ref{Eq25} and \ref{Eq26}}).
         For a description of the dashed line see the text.}
\label{Fig6}
\end{figure}

We start out by describing the $np$ differential cross sections with
the standard one exponential parameterization. Using a restriction on $|t|$ as in
the case of $pp$ scattering  ($|t|<$ 0.125~(GeV/c)$^2$) we found only one data set
\cite{Terrien} at $P_{lab}=$ 0.924 -- 1.793~GeV/c containing the necessary points.
Since that data set was not sufficient for the fit, we increased the upper limit of $|t|$ to
0.25~(GeV/c)$^2$ using the data \cite{Terrien} -- \cite{Arefiev} but
were not satisfied by the fit results because they could not allow to determine
the momentum dependence of the parameters. Various restrictions on the fit range of $t$
were used in the literature \cite{Kreisler,Engler,Perl,Gibbard,Ringia}. Very often
a value for the maximum $|t|$ of 0.5~(GeV/c)$^2$ was considered. Fit results of
Eq.~\ref{Eq1} to the experimental data with this upper limit are shown in Fig.~\ref{Fig6}.

Three problems arise: 1 -- the parameter $A$ at $P_{lab}=$0.924~GeV/c is much larger
than the parameters at similar momenta; 2 -- points at $P_{lab}=$1.97 -- 6.77~GeV/c have large
error bars; 3 -- values of $A$ at $P_{lab}=$23 and 25~GeV/c are lower than the other values.
In addition, the point at $P_{lab}=$23~GeV/c has large error bars, and  there is
a large $\chi^2/NDF=$ 7.82 at $P_{lab}=$9~GeV/c. The problematic points are marked by arrows in
Fig.~\ref{Fig6}.

The large error bars at $P_{lab}=$1.97 -- 6.77~GeV/c are connected with the small number of experimental
points included in the fit, only 4 points for each data set. The same is true at $P_{lab}=$23~GeV/c.
Only 5 data points were considered for that fit. Of course, the number of included data points can
be increased  by increasing the maximum value of the allowed $|t|$ range, but this systematically effects the values of $B$.

\begin{figure}[cbth]
\begin{center}
\includegraphics[width=150mm,height=60mm,clip]{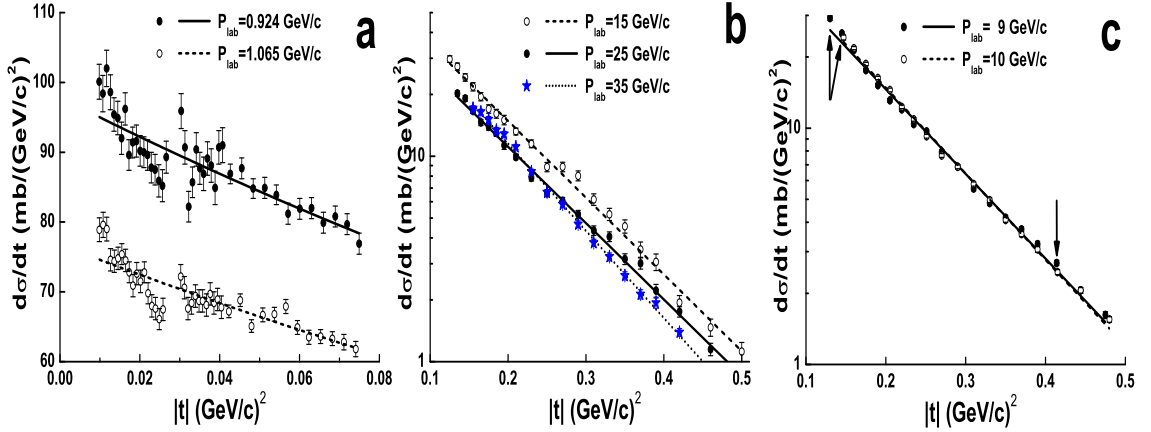}  
\end{center}
\vspace{-5mm}
\caption{ Data points are from Refs. \protect{\cite{Terrien}, \cite{Bohmer}, \cite{Stone}}
 for Figs. a, b and c, respectively. Lines are fit results described in the text.}
\label{Fig7}
\end{figure}

A more complicated situation takes place with other problems (see Fig.~\ref{Fig7}). Fluctuation of
experimental data points \cite{Terrien} at $P_{lab}=$0.924 and 1.065~GeV/c presented in Fig.~\ref{Fig7}a
are comparable to each other. Relative error bars of the data  are also comparable. The only
essential difference is the magnitudes. The difference was noted in Ref.~\cite{Terrien},
but no explanation was given. The data at $P_{lab}=$0.924~GeV/c also are above the phase-shift analysis
results as shown in Ref.~\cite{Terrien}. Thus, we decided to exclude the fit results at $P_{lab}=$0.924~GeV/c from our consideration.

We show experimental data \cite{Bohmer} at $P_{lab}=$15, 25 and 35~GeV/c together with fit results
in Fig.~\ref{Fig7}b. As seen, the slope parameter increases going from 15 to 35~GeV/c. The $B$ values
at 15 and 25~GeV/c are comparable.  At the same time the maximum value of $d\sigma/dt$
(at $|t|<$ 0.23~(GeV/c)$^2$) decreases going from 15 to 25~GeV/c, and suddenly increases going to
35~GeV/c.  This behavior was not noted in Ref.~\cite{Bohmer}. Maybe it was not
essential because the absolute normalization error was estimated to be $\sim$ 35\% \cite{Bohmer}.
We did not consider systematic errors in our fit. Because the fit results at $P_{lab}=$25~GeV/c
fall outside of the common trend, we do not take them into account.

We show in Fig.~\ref{Fig7}c experimental data \cite{Stone} at two similar momenta, 9 and 10~GeV/c together with
our fit results. The data are similar, except data points at $|t|=$0.13 and 0.145~(GeV/c)$^2$
at $P_{lab}=$9~GeV/c (marked by arrows). They give the largest contribution to $\chi^2$. The other
similar data point is also marked in the figure. Because the general properties of the distributions
are similar, we consider the large $\chi^2/NDF$ at $P_{lab}=$ 9~GeV/c to be a consequence of the data
quality and omission of the systematic errors.

The fit results are connected with the fundamental properties of $np$ interactions --
$\sigma^{tot}_{np}$, $\sigma^{el}_{np}$ and $\rho_{np}$. According to Eq.~\ref{Eq3}, $A$ can be
calculated as $A=\sigma^{tot}_{np}\sqrt{1+\rho_{np}^2}/(4\sqrt{\pi})$. The calculations with the
assumption $\rho_{np}=\rho_{pp}$ are shown in Fig.~\ref{Fig6}a by the dashed line. As seen,
the calculations deviate from the fit results, especially for $P_{lab}\leq$ 10~GeV/c. We used
Eq.~\ref{Eq23} as an approximation for $\sigma^{tot}_{np}$, and Eq.~\ref{Eq6} for $\rho_{np}$.
At the same time, having $\sigma^{tot}_{np}$ we can calculate $\rho_{np}$ using Eq.~\ref{Eq3}
and the following approximation for $A$:
\begin{equation}
A=2.8+6\ \frac{\sqrt{s/s_0-4\cdot 1.072}}{(s/s_0-5)^2+20} \ \ \ [{\rm fm/(GeV/c)}].\label{Eq25}
\end{equation}
The calculations of $\rho_{np}$ are shown by dashed lines in Fig.~\ref{Fig5}, and is seen to differs from
$\rho_{pp}$. Consequently, our results show that a standard assumption,
$\rho_{np}=\rho_{pp}$, is not correct.

Having $A$ and $B$ given by Eqs.~\ref{Eq25} and \ref{Eq26}, we can estimate $\sigma^{el}_{np}$ as
$A^2/B$. In doing this $B$ is approximated by:
\begin{equation}
B=6.2 + 0.7\ln(s/s_0)-\frac{350}{(s/s_0)^3} \ \ \ [({\rm GeV/c})^{-2}]. \label{Eq26}
\end{equation}
The estimations of $\sigma^{el}_{np}$ are shown by the dashed line in Fig.~\ref{Fig5}.
As seen, they underestimate the cross section at $P_{lab}\leq$ 5~GeV/c. This was expected
because the standard parameterization cannot describe the cross sections at large momentum
transfer. Thus, we consider the two exponential parameterization.

\begin{figure}[cbth]
\begin{center}
\includegraphics[width=150mm,height=45mm,clip]{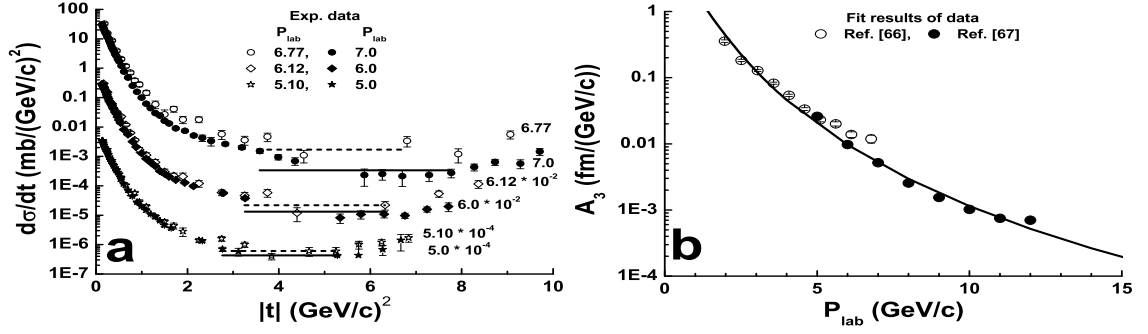}
\end{center}
\vspace{-5mm}
\caption{a) Differential cross section distributions of $np$ elastic scattering at $P_{lab}=$ 5.1, 6.12 and
         6.77~GeV/c \protect{\cite{Perl}} (open points) and at $P_{lab}=$ 5, 6 and 7~GeV/c
         \protect{\cite{Stone}} (closed points). Dashed and solid lines are fit results
         by Eq.~\ref{Eq27} with $A_1=A_2=0$ to the data \protect{\cite{Perl}}
         and \protect{\cite{Stone}}, correspondingly. b) $A_3$ values from the fits to the data.
         The solid line is Eq. \ref{Eq28}.}
\label{Fig8}
\end{figure}

Typical differential cross section distributions of $np$ elastic scattering are shown in Fig.~\ref{Fig8}a
and have two maxima at forward and backward directions in the center-of-mass reference frame, at
$t\sim 0$ and $t\sim t_{max}$. The backward peak appears due to the charge exchange
reaction, $n+p\rightarrow p+n$. It is assumed that the backward peak is connected with $\pi$-meson
exchange in the $t$ channel (see references in \cite{Friedes}). The backward peak is located at
$|u|<$ 0.025 (GeV/c)$^2$ ($u=t_{max}-t$), and it is much smaller than the forward peak.
Thus, we will not consider it in the following.

The forward peak is located at $|t|\leq$ 0.5~(GeV/c)$^2$ \cite{Stone}. There is a change of slope
at $|t|\sim$ 1.5 (GeV/c)$^2$, and a plateau at
$0.3\cdot |t_{max}| \leq |t| \leq 0.7\cdot |t_{max}|$. It is obvious that the two exponential
parameterization cannot describe the plateau. Thus, we modify the parameterization to be:
\begin{equation}
\frac{d\sigma}{dt}=|A_1\ e^{B_1t/2}\ + e^{i\phi}(A_2\ e^{B_2t/2}+A_3)|^2.\label{Eq27}
\end{equation}

A fit of Eq.~\ref{Eq27} to the experimental data of Refs.~\cite{Perl,Stone} at
$0.3\cdot |t_{max}| \leq |t| \leq 0.7\cdot |t_{max}|$ and $A_1=A_2=0$ shows that the height of
the plateau decreases as the momentum increases, and can be described by:
\begin{equation}
A_3=2000/(s/s_0)^{4.75}\ \ \ [\rm fm/(GeV/c)]. \label{Eq28}
\end{equation}

The plateau corresponds to isotropic scattering in the center-of-mass reference frame.
Its yield in the differential cross section distributions are shown in Fig.~\ref{Fig8}a by solid and
dashed lines.

As seen in Fig.~\ref{Fig8}b, the fit results to the data \cite{Perl,Stone} differ.
To understand the source of the difference we plotted the differential
cross sections of Refs.~\cite{Perl,Stone} at similar projectile momenta in Fig.~\ref{Fig8}a.
As seen, the data
of Ref.~\cite{Perl} are less precise than the data of Ref.~\cite{Stone} in the region of
the plateau. The data in \cite{Perl} vary more stronger than the data in \cite{Stone}.
The data are quite close to each other only at $P_{lab}\sim$ 5~GeV/c. Thus, we  mainly
followed  the data of Ref.~\cite{Stone} to fulfill Eq.~\ref{Eq28},
shown in Fig.~\ref{Fig8}b by a solid line.

\begin{figure}[cbth]
\begin{center}
\includegraphics[width=150mm,height=90mm,clip]{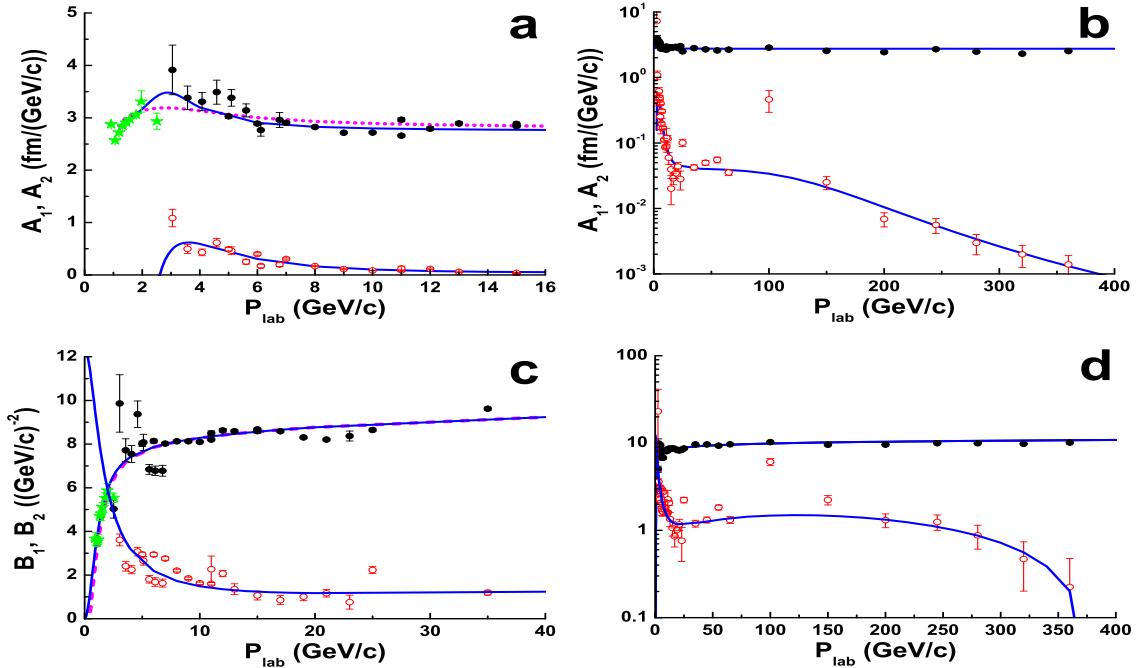}
\end{center}
\vspace{-5mm}
\caption{Fit results of Eq.~\ref{Eq27} to the data \protect{\cite{Perl}--\cite{DeHaven}}.
         Points are the fit results. Stars (green) are the results of fitting of Eq.~\ref{Eq27}
         with $A_2=0$ to the data \protect{\cite{Terrien}}. Solid lines are approximations
         (see below). Dashed lines are approximations of the $A$ and $B$ parameters of the one
         exponential expression.
}
\label{Fig9}
\end{figure}

A two exponential fit of Eq.~\ref{Eq27} to the data \cite{Terrien,Perl,Stone,Bohmer,DeHaven}
using Eq.~\ref{Eq28} gives meaningful results only for 22 of 45 sets of data. This is a consequence of the strong correlation of the parameters. According to the fit, an average value of
$\phi$ is equal to 1.6. In order to reduce the correlations, we fixed $\phi$ to that value. With that constraint 39 data sets could be included. The fit results are shown in Fig.~\ref{Fig9}.

The fit to the data of Ref.~\cite{Terrien} at $P_{lab}=$ 0.924 -- 1.793~GeV/c gives too
large values of $A_1$ and $B_1$, which are needed to reproduce the data at
$|t|<$ 0.027~(GeV/c)$^2$ (see Fig.~\ref{Fig7}a). The large values of $A_2$ and $B_2$ of the fit allow
 the data at $|t|>$ 0.03~(GeV/c)$^2$ to be described. The errors of the parameters are also large.
As seen in Fig.~\ref{Fig7}a, there is an empty region at 0.027 $<|t|<$ 0.03~(GeV/c)$^2$.
We believe that the region is a reflection of special features of the experiment, which also
leads to a difference between the cross sections before and after the region. Taking all of
these points into account, we conclude that the fit results are not realistic and thus we do not show them
in Fig.~\ref{Fig9}.

At the same time, the data \cite{Terrien} are fit quite well by Eq.~\ref{Eq27} with
$A_2=0$ and $A_3$ given by Eq.~\ref{Eq28}. Those  results are indicated by stars in Fig.~\ref{Fig9}.
The parameters $A_1$ and $B_1$ in this case are rather close to the results of the one
exponential parameterization fit.

\begin{figure}[cbth]
\begin{center}
\includegraphics[width=150mm,height=100mm,clip]{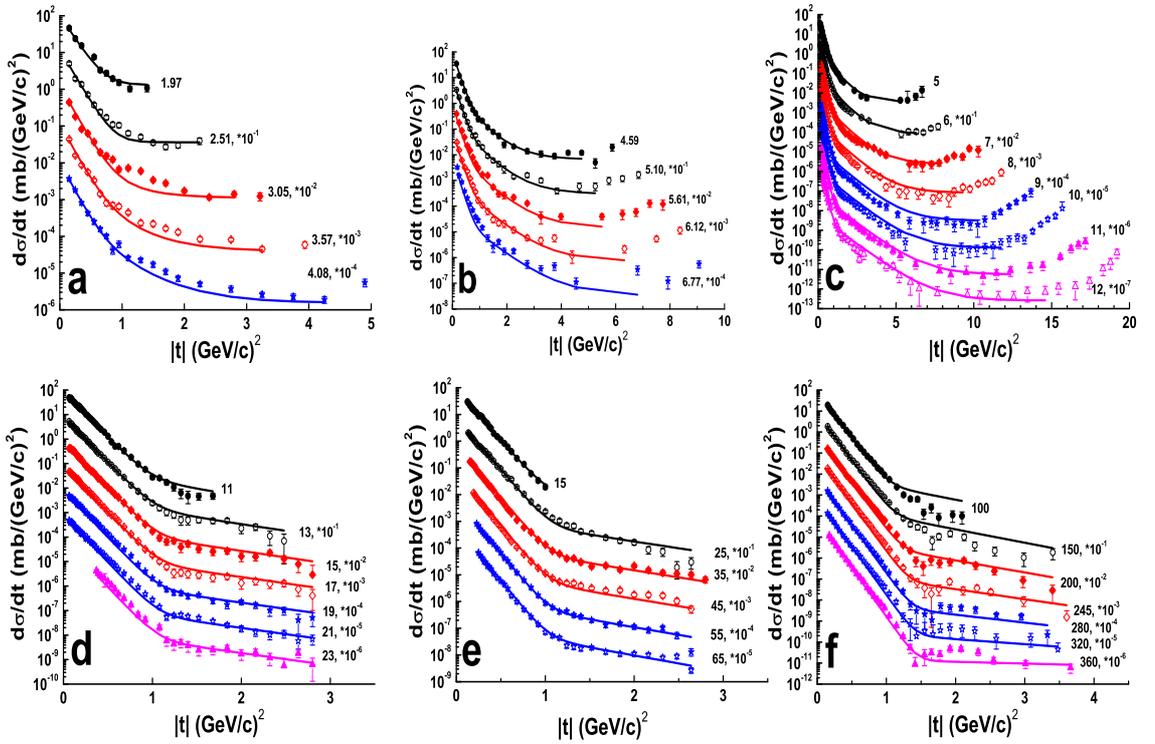}
\end{center}
\vspace{-5mm}
\caption{Description of $np$ elastic differential cross sections by Eq.~\ref{Eq27} with
         approximations \ref{Eq29} -- \ref{Eq32}. Points are experimental data
         \protect{\cite{Perl,Stone,Bohmer,DeHaven}}. The experimental errors displayed are statistical
         only. Lines are calculation results.
}
\label{Fig10}
\end{figure}

The data of Ref.~\cite{Perl} at $P_{lab}=$ 1.97 and 2.51 GeV/c do not have sufficient points at large
$|t|$ (see Fig.~\ref{Fig10}a) for a good determination of the parameters. Thus, we fit Eq.~\ref{Eq27}
with $A_2=0$ to them as before. The fit results are also shown in Fig.~\ref{Fig9} by stars.

The fit results of the data \cite{Perl} at $P_{lab}=$3.05, 3.57, 4.08, 4.59, 5.10, 6.12 and
6.77~GeV/c  have large error bars. The results are much better for the data \cite{Stone} at
$P_{lab}=$5, 6, 7, 8, 9, 10, 11 and 12~GeV/c.

The data of Ref.~\cite{Bohmer} at $P_{lab}=15$~GeV/c could not be fit at all due to the restricted
range of $t$ (see Fig.~\ref{Fig10}e). The peculiarity of the data \cite{Bohmer} at
$P_{lab}=25$~GeV/c was considered before. They lead to $A_2$ and $B_2$ values that are not consistent with other data sets.

The restricted range of $t$ reflected on the fit results of data \cite{DeHaven} at $P_{lab}=100$~GeV/c
(see Fig.~\ref{Fig10}f). At higher momenta all parameters regularly decrease with encreasing momentum,
up to $P_{lab}=360$~GeV/c.

Taking into account everything above, we propose the following approximation for the momentum dependence
of the parameters at $P_{lab} \geq 2.5$~GeV/c:
\begin{equation}
A_1=2.75+ 2.25\ \frac{\sqrt{s/s_0-4.3}}{(s/s_0-7)^2+5}-1.4\ 10^7/(s/s_0)^{12}\ \ \ [\rm fm/(GeV/c)], \label{Eq29}
\end{equation}
\begin{equation}
A_2=1.7\ 10^4\ (s/s_0-7)/(s/s_0)^5+0.04/[1.4\ 10^{-10}(s/s_0)^4+1] \ \ \ [\rm fm/(GeV/c)], \label{Eq30}
\end{equation}
\begin{equation}
B_1=\ 6.2+0.7\ \ln{(s/s_0)}-310/(s/s_0)^3\ \ \ [(GeV/c)^{-2}], \label{Eq31}
\end{equation}
\begin{equation}
B_2=2\ 10^{-4}\sqrt{s/s_0}\ (715-s/s_0)+80/(s/s_0)^{1.5}\ \ \ [(GeV/c)^{-2}]. \label{Eq32}
\end{equation}
These parameters allow  general features of the differential cross sections of $np$ elastic
scattering starting from 400~MeV up to 360~GeV to be described. The description is shown in
Fig.~\ref{Fig10}. The total $\chi^2/NDF$ for all data sets is equal 14740/1290$\sim$11 without
considering the systematic uncertainties since these are either not consistently provided, or in some cases not at all. For example, 4--7 \%
uncertainty in absolute normalization is given in Ref. \cite{Terrien}, 10--20 \% --
in Ref. \cite{Perl},  +5 -- -15 \% -- in Ref. \cite{DeHaven}. We have used average values in these
cases. Taking into account such uncertainties we have obtained $\chi^2/NDF= 4938/979 \sim 5$,
which is vastly improved.\footnote{We did not include in the fit the data of Ref. \cite{Engler}
because the systematic uncertainty was not presented in the paper.}

The worst $\chi^2/NDF$ is observed for the data \cite{DeHaven} at $P_{lab}\geq$ 100~GeV/c.
A separate fit of the data
gives an acceptable $\chi^2/NDF$, $A_1$ and $B_1$ which deviate from the corresponding
approximations in the range by $\pm$ 5 \%. Fitted values of $A_2$ and $B_2$ vary in a larger
interval. The uncertainty of the $A_2$ values is $\sim$ 25 \%, and errors of $B_2$ are on the level
$\sim$ 50 -- 70 \%. This is understandable because the data do not include points with large
momentum transfer. Thus, $A_2$ and $B_2$ cannot be determined as well in that momentum range.
At the same time, our approximations allow the forward peak to be described sufficiently well.

\section*{Conclusions}
\begin{itemize}
\item
A general description of $pp$ and $np$ elastic scattering in the beam momentum range 2 -- 50~GeV/c
has been reached.

\item
133 and 45 sets of experimental data on $pp$ and $np$ elastic scattering, respectively, were
analyzed and fit.

\item
Two popular parameterizations of differential cross sections -- a standard one exponential
parameterization and the two coherent exponential parameterization, were used to fit
the experimental data.

\item
Analytical expressions to approximate the momentum dependence of the fit parameters were proposed.

\item
Approximations of $\sigma^{tot}$, $\sigma^{el}$ and $\rho$ have also been proposed.

\end{itemize}

All of these give a solid base for effective Glauber calculations and Monte Carlo simulations
of properties of nucleon-nucleon, nucleon-nucleus and nucleus-nucleus interactions at high momenta
especially for FAIR and NICA, and for the RHIC Beam Energy Scan program.

The authors are thankful to D. Mchedlishvili for providing us with COSY data in a tabulated form,
and O.V. Selyugin and N.I. Kochelev for useful discussions of the subject of the paper.

\end{document}